\documentclass[aps,prl,twocolumn,superscriptaddress]{revtex4}
\usepackage{amsmath,bm}
\usepackage{mathrsfs}
\usepackage{amsfonts}

\usepackage{graphicx}
\usepackage{ulem}

\usepackage{graphicx}
\usepackage{epstopdf}
\usepackage{dcolumn}
\usepackage{amsmath}
\usepackage{epsfig}
\usepackage{indentfirst}
\usepackage{psfrag}
\usepackage{subfigure}
\usepackage{amssymb}
\usepackage{color}

\usepackage{graphicx}
\usepackage{dcolumn}
\usepackage{bm}
\usepackage{natbib}

\makeatletter
\newcommand\colorsout[1]{\bgroup \markoverwith{\textcolor{#1}{\rule[0.5ex]{2pt}{0.4pt}}}\ULon}

\makeatother

\begin{document}

\title{Mapping the orbital structure of impurity bound states in a superconductor}

\author{D.-J. Choi}
\affiliation{CIC nanoGUNE, Tolosa
Hiribidea 76, 20018 San Sebasti\'an - Donostia, Spain}

\author{C. Rubio-Verd\'u}
\affiliation{CIC nanoGUNE, Tolosa
Hiribidea 76, 20018 San Sebasti\'an - Donostia, Spain}

\author{J. de Bruijckere}
\affiliation{CIC nanoGUNE, Tolosa
	Hiribidea 76, 20018 San Sebasti\'an - Donostia, Spain}
\affiliation{Kavli Institute of Nanoscience, Delft University of Technology, Lorentzweg 1, 2628 CJ Delft, The Netherlands}

\author{M.M. Ugeda}
\affiliation{CIC nanoGUNE, Tolosa
	Hiribidea 76, 20018 San Sebasti\'an - Donostia, Spain}\affiliation{ikerbasque,
	Basque Foundation for Science, Bilbao, Spain}

\author{N. Lorente}
\affiliation{Centro de F\'{\i}sica de Materiales CFM-CSIC, 20018 San Sebasti\'an - Donostia, Spain}
\affiliation{Donostia International Physics Center (DIPC), Paseo Manuel de Lardizabal 4, 20018 San Sebasti\'an-Donostia, Spain}

\author{J.I. Pascual}
\affiliation{CIC nanoGUNE, Tolosa
Hiribidea 76, 20018 San Sebasti\'an - Donostia, Spain} \affiliation{ikerbasque,
Basque Foundation for Science, Bilbao, Spain}

\date{\today}

\begin{abstract}

  A magnetic atomic impurity inside a superconductor locally distorts  superconductivity. They scatter Cooper pairs as a potential with  broken time-reversal symmetry, what leads to localized  bound states with subgap excitation energies, named hereon Shiba states. Most conventional approaches to study Shiba states  treat magnetic impurities as point scatterers with an isotropic exchange interaction, while the complex internal structure of magnetic impurities is usually neglected. Here, we show  that the number and the shape of Shiba states are correlated to the spin-polarized atomic orbitals of the impurity, hybridized with the superconducting host, as supported by Density Functional Theory simulations.	 
  Using  high-resolution scanning tunneling spectroscopy, we spatially map the five Shiba excitations found on sub-surface chromium atoms in Pb(111), resolving both their particle and hole components.  While the maps of particle components  resemble  the   \textit{d} orbitals of embedded Cr atoms, the  hole components  differ strongly from them. 
  The orbital fingerprints of Shiba states thus unveil the magnetic ground state of the impurity, and identify scattering channels and interactions, all valuable tools for  designing atomic-scale  superconducting devices.

\end{abstract}

\maketitle

Yu-Shiba-Rusinov (Shiba)  states \cite{Yu_1965,Shiba_1968,Rusinov_1969} are identified in scanning tunneling spectra as pairs of intra-gap resonances symmetrically positioned around  zero-bias  \cite{Yazdani_1997,Hudson2001,Ji_2008,Franke_2011,Ruby_2015,Gerbold_2015}. Each resonance is the excitation of  an electron or hole   into the bound state \cite{Ruby_2015}, thus representing the particle or hole components of the  quasiparticle  wavefunction.  Since Shiba excitations lie inside the superconducting gap, their lifetime considerably exceed that of other quasiparticles (QPs). This anticipates that Shiba peaks behave as a robust probe of scattering phenomena in superconductors, revealing intrinsic properties  such as the distribution of the order parameter \cite{Hudson2001}, the QP band structure \cite{McElroy2003}, the effect of dimensionality \cite{Gerbold_2015}, or Andreev tunneling processes \cite{Ruby_2015}. Owing to their long lifetime, Shiba excitations exhibit narrow lineshapes, which enables the study of magnetic phenomena in impurities with high energy resolution, such as  magnetic anisotropy \cite{Zitko2011,Hatter_2015} and magnetic coupling \cite{Ji_2008,Yao2014}. 

\begin{figure}
	\hfill\includegraphics[width=1.\columnwidth]{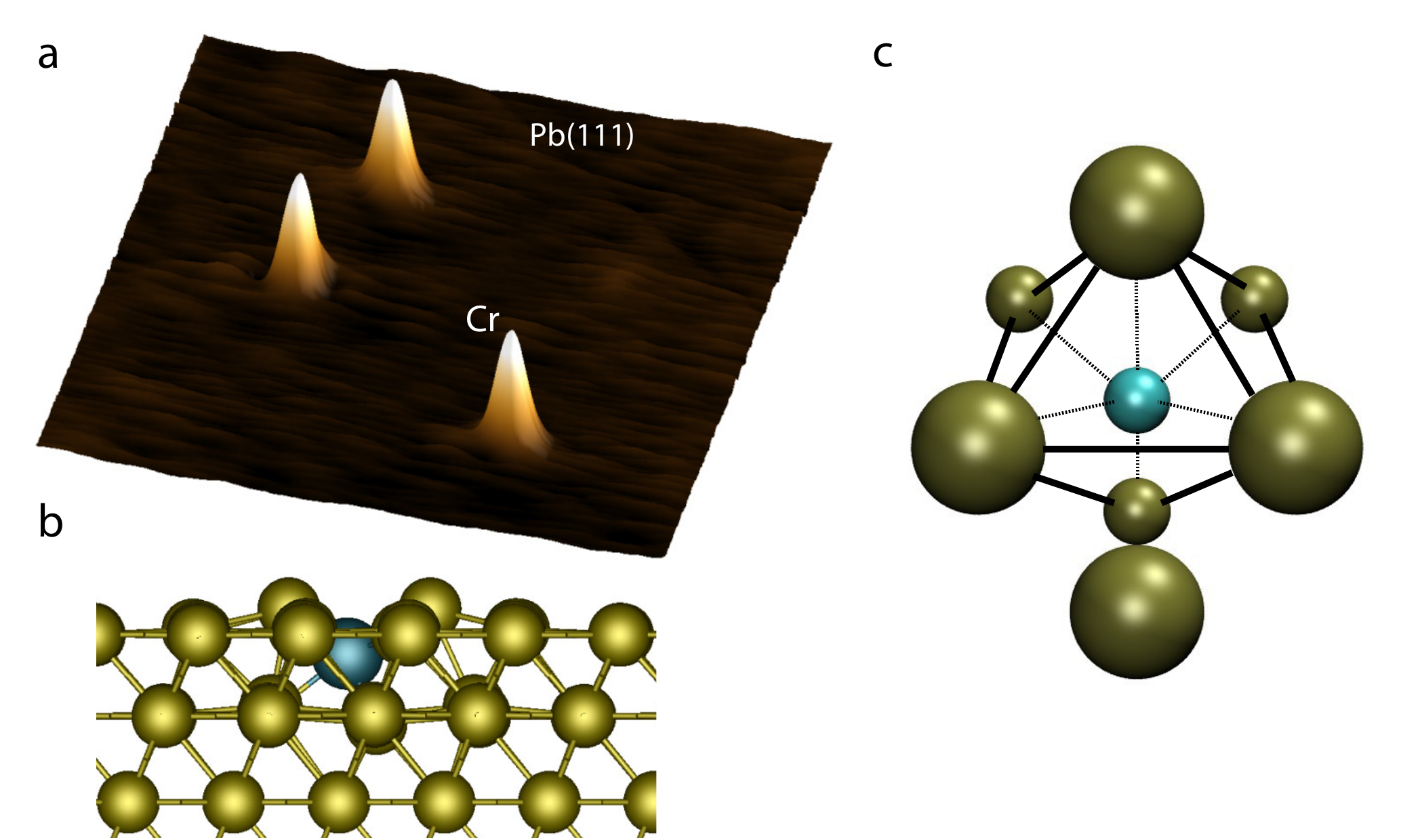}
	\caption{\label{Figure1} 
		Cr impurities in Pb (111). $(a)$ Constant-current STM image of
		three Cr atoms in the Pb (111) surface (V=4 mV, I=10 pA, $12\times12$ nm$^2$). The atom apparent height is 50 $pm$ for these imaging conditions. 
		$(b)$ Results of a full DFT relaxation showing that the most stable position of a Cr atom (blue)  is at a sub-surface site (Pb atoms in yellow).      $(c)$ Top view of the local atomic neighborhood of the embedded Cr atom, composed by seven Pb atoms all at a  distance  of about 3  $\pm 0.2$  \AA.   The four largest Pb atoms are surface atoms. The relaxed Cr atom lies in a  pseudo-FCC position,   displaced towards two Pb atoms, which are slightly lifted above the surface plane. The resulting cluster resembles a distorted octahedron with an extra atom at one vertex.	}
\end{figure}

\begin{figure*}
	\includegraphics[width=0.9\textwidth]{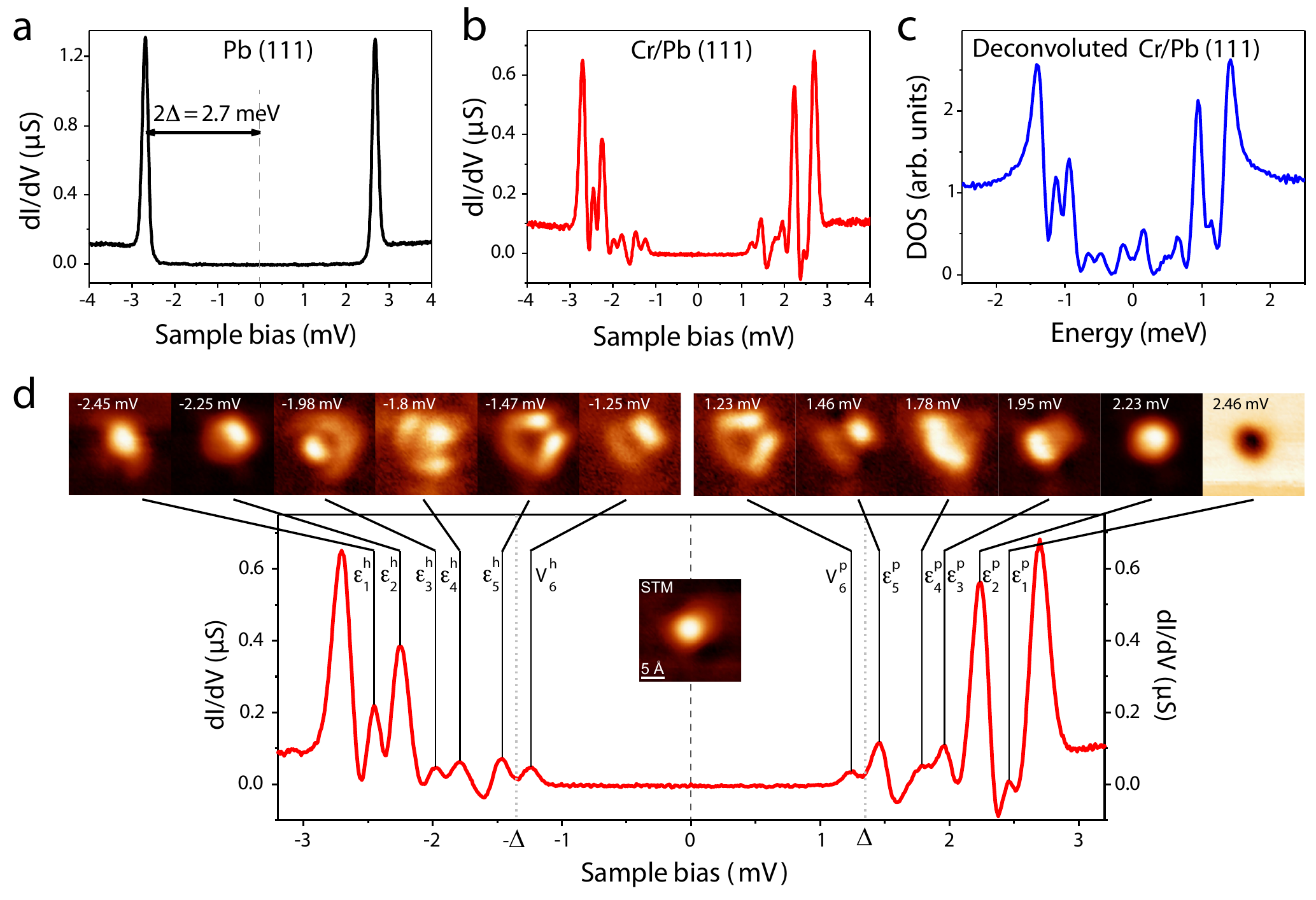}
	\caption{\label{Figure2}
		Differential conductance spectra measured with a superconducting (Pb-covered~\cite{Franke_2011,Ruby_2015c}) tip $(a)$ on the bare Pb(111)  surface and $(b)$ on the Cr adatom,  where intra-gap states are easily recognized.  $(c)$ DOS of the Cr impurity obtained by deconvoluting the tip contribution  in  $(a)$  from spectrum  $(b)$, as described in  the SI \cite{SOM}. Five positive- and five negative-bias states are observed inside the superconducting gap, at opposed bias-polarity  values, in agreement with five pairs of Shiba excitations with different  intensities. Most of the differences in peak intensity are due to their inhomogeneous spatial distribution \cite{SOM}. $(d)$ dI/dV maps over the Cr impurity  at the energy of the intra-gap states in the spectrum  $(b)$ (expanded here for clarity). The maps represent the amplitude of each spectral peak obtained from a 52$\times$52  matrix of dI/dV spectra  acquired in an area of 2.2 nm$\times$2.2 nm over the chromium atom (all with set point V=4 mV and I=0.35 nA). Gap peaks are labeled according to their position and hole(h)/particle(p) character. Peaks $v_6^h$/$v_6^p$ correspond, respectively, to injection of holes/electrons into states  $\epsilon_5^p$/$\epsilon_5^h$, which are thermally populated by particles/holes due to their proximity to E$_F$. Therefore, states $\epsilon_5^p$ and $\epsilon_5^h$ are imaged twice. 
	}
\end{figure*}

Theoretical models using  classical spins with isotropic exchange fields have predicted multiple Shiba bound states associated to angular momentum quantum numbers \cite{Rusinov_1969}. However, magnetic transition metal (TM) atoms in a  superconductor show a varying number of Shiba bound states, that depends on the nature of both the  element and  the superconductor \cite{Pan2000a,Hudson2001,Ji_2008,Ruby_2015,Gerbold_2015}. Due to the orbital character  of the  scattering channels of TM impurities, it is expected that Shiba multiplets reflect the occupation level of the atomic shell \cite{Moca_2008}, what would render them as the ideal probe for identifying the magnetic ground state of a single impurity in a superconductor. However, the orbital character of Shiba states remains elusive in the scarce previous attempts to experimentally map them \cite{Hudson2001,Ji_2008,Gerbold_2015,Randeria2016}, while their spatial extension has been mostly associated with the properties of the superconducting host. 

Here we explore  possible orbital components in Shiba states by comparing the QP local density of states (LDOS) of chromium atoms deposited on Pb(111), measured with a low temperature scanning tunneling microscope (STM),  with atomistic simulations based on Density Functional Theory (DFT). We deposited  Cr atoms  on a Pb(111) film grown on SiC(0001) (thickness $>$ 100 nm) at 15 K. The atoms appear in STM images (measured at T = 1.2 K) as protrusions with a small apparent height of $\sim 50\,$ pm (Fig.~\ref{Figure1}~a) \cite{Ruby_2015c}, suggesting a subsurface atomic configuration of the Cr atoms. This is further supported by our DFT calculations (see Methods section and Supplementary Information (SI) for details), which reveal that the most stable configuration corresponds to the Cr atom underneath the uppermost Pb layer, as sketched in Figs. 1b and 1c. 

\begin{figure*}
 \includegraphics[width=0.9\textwidth]{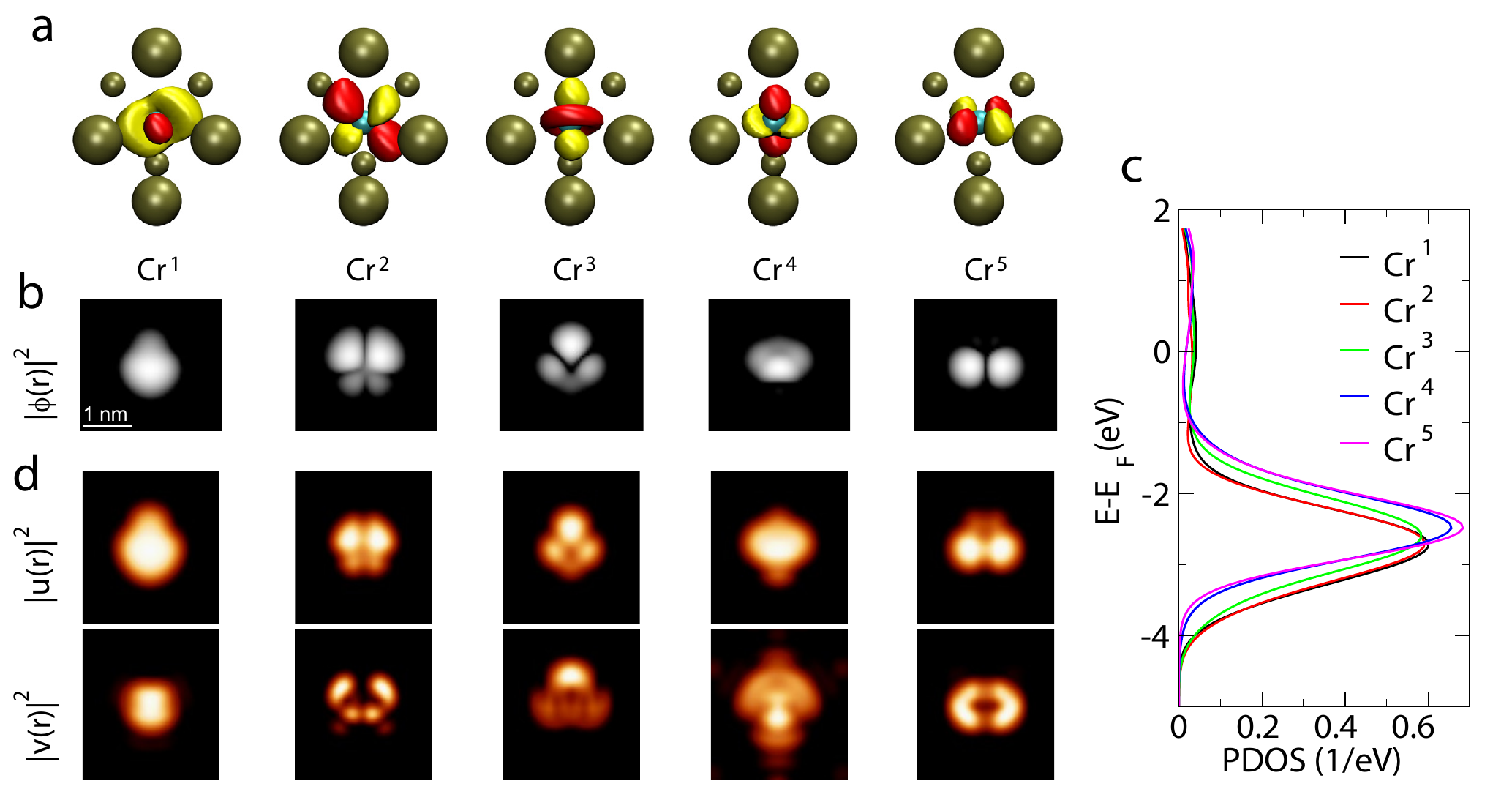}
	\caption{\label{Figure3}
	  DFT results of $d$ scattering states of  Cr/Pb(111)  and their corresponding  Shiba states. $(a)$  Top-view of wavefunction amplitude isosurfaces for the five electronic states of the cluster shown in Fig. 1c. These five states  are the only ones with relevant weight from the original $d$-manifold of the Cr atom, and their shape resemble  the Cr $d$-electrons modified by the symmetry of the embedded  site,  as can be seen in  the top view of their squared amplitude modulus in  $(b)$.
	  $(c)$ Spin-polarized projected density of states (PDOS) of the full  electronic structure of the Pb(111)+Cr slab (Fig. 1b)  on the five cluster orbitals depicted in $(a)$. The cluster  states are considerably broadened due to the strong coupling  with the complete Pb system, and spin-polarized: only one spin component is shown here,  the other component shows only PDOS well above the Fermi level. 
	   $(d)$  Modulus squared of the Bogoliubov quasiparticle coefficients,  i.e. $|u_n (\vec{r})|^2$ and $|v_n (\vec{r})|^2$,  on the Pb(111) surface, calculated for the five cluster states of $(a)$ (see SI \cite{SOM}).   }
\end{figure*}

The quasiparticle LDOS associated to subsurface Cr impurities was obtained from differential tunneling conductance (dI/dV) spectra using  superconducting Pb-terminated tips to increase the energy resolution beyond the thermal limit \cite{Franke_2011}. Figure 2a shows a typical  dI/dV spectrum acquired on bare Pb(111) that exhibits a doubled SC gap of Pb with sharp coherence peaks at $\pm$ 2$\Delta = 2.7$ meV \cite{note1}.  The dI/dV curve taken on a Cr atom in Fig. 2b reveals a rich spectral structure inside the superconducting gap, with six intra-gap peaks at each polarity (i.e. six peak pairs), in addition to the original Pb coherence peaks. To obtain the quasiparticle LDOS, we deconvoluted the experimental dI/dV spectrum on chromium (Fig. 2b) using the superconducting  DOS of a Pb-tip, estimated from Fig. 2a (details in SI \cite{SOM}). The deconvoluted spectrum, shown in Fig. 2c, shows just five energy-symmetric pairs of peaks,  labeled as $\epsilon_n^{h,p}$ (n=1, 2, 3, 4, 5) in Fig. 2d. We attribute these pairs of peaks to five Shiba states with excitation energies    $\epsilon_1=\pm 1.1$ meV, $\epsilon_2=\pm 0.9$ meV, $\epsilon_3=\pm 0.62$ meV, $\epsilon_4=\pm 0.45$ meV and $\epsilon_5=\pm 0.125$ meV.  An additional pair of peaks, closest to zero bias in the  dI/dV spectrum of Fig. 2b, denoted V$_6^h$ and V$_6^p$, appear at bias voltages $\pm(\Delta-\epsilon_5)/e$ and, thus, corresponds to a thermal replica  of the Shiba pair  $\epsilon_5^h$/$\epsilon_5^p$ \cite{Ruby_2015}. Shiba multiplets arise naturally as different angular momentum components in isotropic exchange fields. However, the resolution of a five-fold Shiba multiplet  achieved here enables us to reinterpret their origin in terms of atomic properties of the impurity  \cite{Moca_2008,Zitko2011}. 

To probe the orbital origin of the Shiba bound states, we map their intensity in a small region around the Cr impurity. Figure 2d shows dI/dV maps obtained for the same Cr atom at the voltages of all twelve peaks found inside the SC gap (See Methods).
In contrast to the featureless protrusion in topography (inset of Fig. 2d), 
the conductance maps reveal various features  around the  Cr impurity. 
Each Shiba state appears with a different shape  within the same polarity, as well as at the corresponding opposite polarity. 
For example, the shape of the QP states $\epsilon_3^p$ and $\epsilon_4^p$, and $\epsilon_5^p$ are markedly different from their corresponding hole states, which reveals a particle-hole asymmetry in the  wavefunction of the corresponding Shiba states. 
Such spatial asymmetry is also observed in the thermal replicas at $V_6^h=-(\Delta-\epsilon_5^p)/e$ and $V_6^p=(\Delta-\epsilon_5^h)/e$, 
which show the shapes of their mirror states $\epsilon_5^p$ and $\epsilon_5^h$, respectively.  The different shape of particle and hole components of every state explains their different  peak amplitude in  point spectra like in Fig. 2b. In fact, dI/dV spectra averaged over the whole  extension around the Cr impurity levels their intensity at both polarities (see SI \cite{SOM}).

To find out the origin of the number and shape of the Shiba states, we simulated with DFT (see methods) the scattering channels of the embedded Cr impurity and modeled their effect in producing Shiba bound states. The channels are very sensitive to atomic-scale details of the Cr atom and its environment. We employed the most stable Cr site in the subsurface plane described in Figure 1b (see SI ~\cite{SOM}),  and found that this site produced  results compatible with the experiment. 

The identification of Cr-derived scattering channels is
difficult; the embedded atom does not lie in a fully symmetric position inside
the Pb FCC crystal, while it induces a significant displacement
on its  first Pb neighbors.   Consequently, the Cr $d$ orbitals
are strongly hybridized and mixed with Pb $sp$ bands. To facilitate
their identification in the band's continuum, we calculated first the
single-electron energy levels of the cluster formed by the Cr atom and the
7 first neighbours with apparent interaction (shown in Fig.~1c). We found that, at
least in this reduced  ensemble, there are five singly-occupied states with clear
$d$ character centered in the Cr atom. Figure 3a plots isosurfaces of wavefunction amplitude for the five eigenstates inside the atomic cluster. Their shape partly resembles   the  original symmetry of $d$ orbitals. In the full Pb system, these states appear as broad resonances around -2.6 eV  (Fig.~3c shows the  DOS of the Pb slab projected on the five cluster states of Fig. 3a)   due to the mixture with $sp$ bands of the lead film, but their spin polarization remains very strong. We deem that these five one-electron states  host  the five spin-polarized scattering channels, through which the intra-gap bound states are created.

The spatial shapes  of the  five Cr-states of  the  cluster,  projected on the surface  (Fig. 3b), show already some level of agreement with the experimental Shiba maps. For example, the maps of peaks $\epsilon_4^h$,   $\epsilon_5^p$, or  $\epsilon_3^p$  resemble the shapes of states Cr$^3$, Cr$^2$, or Cr$^4$, respectively. 
To  correlate orbital states  with subgap impurity bound states, we computed the corresponding  Shiba states by considering that  the effect of an impurity on the total Hamiltonian can be divided in potential (no spin degrees of freedom) and in spin contributions \cite{Rusinov_1969}, such that

\begin{equation}
\hat{H}_{Imp}= K (\vec{r}) - J (\vec{r}) \vec{S} \cdot \vec{\sigma}.
\label{Ham}
\end{equation}  

In a first approximation,  the five  scattering channels pictured in Fig. 3a diagonalizes these contributions. They are then responsible for the appearance of  five positive-energy (particle)  and five negative-energy (hole) Shiba states in the QP LDOS, symmetrically aligned with respect to the Fermi energy \cite{Moca_2008}. The DFT results indicate that for this system the potential scattering term $K$ in Eq.~(\ref{Ham}) is very small. In the absence of a  potential scattering term, a large degree of particle-hole symmetry in the weight and amplitude of the Shiba states  is expected~\cite{Hudson2001,Flatte_2000}.  This is indeed the case after adding up all differential conductance spectra measured over the surface region around the Cr impurity, as commented above and shown in Fig.~S3 in the SI \cite{SOM}.  

The spatial distribution of Shiba amplitudes can  be depicted by the squared modulus   of the Bogoliubov quasiparticle coefficients, $|u_n (\vec{r})|^2$ and $|v_n (\vec{r})|^2$, representing the  LDOS of their particle  and hole components, respectively~\cite{Balatsky_PRB_2008}. Figure~\ref{Figure3}d plots the quantities $|u_n (\vec{r})|^2$  and $|v_n (\vec{r})|^2$ over the embedded Cr impurity, obtained as described in the SI \cite{SOM}. The results  reproduce the different shape of particle and hole states from the experiments: the particle component   $|u_n|^2$  resembles the shape of Cr $d$ states  in Fig. 3c, while $|v_n |^2$  deviates strongly. This is due to the dominating  spin contribution to the potential: QP scattering with hole components produces a phase reversal for wavevectors of the  Fermi  surfaces of lead, resulting in a clear distortion of the shape of scattering channels \cite{SOM}. 

We can recognize several features from the experimental conductance maps of Shiba states (Fig. 2d) in the simulated particle and hole components of  Bogoliubov quasiparticles (Fig. 3d). In particular,  the particle components $|u_n|^2$ of orbitals Cr$^2$ and Cr$^4$  resemble the shape of  positive bias Shiba maps $\epsilon_5^p$ and  $\epsilon_3^p$, respectively, whereas their  $|v_n |^2$ fits well with the corresponding negative-bias maps, $\epsilon_5^h$ and  $\epsilon_3^h$. However, the opposite behaviour is found for the Shiba bound state $\epsilon_4$: the shape of the particle component matches with the negative bias map. The reversal of bias polarity of particle and hole components is a fingerprint of a quantum phase transition driving a spin channel from a doublet into a Shiba singlet state. This  occurs  when the exchange coupling constant $J$ is large to induce the breaking of a Cooper pair  \cite{Salkola_1997,Franke_2011,Hatter_2015}. In our case, this reveals that the given scattering channel is fully screened into a singlet state due to a larger exchange potential  and, thus, does not contribute to the total atomic spin.   

The sharp maps of Shiba states  change shape rapidly between peaks,  in barely less than 200 $\mu$eV, a much greater accuracy than any other method of orbital imaging one-electron states, and in spite of the large degree of hybridization of the buried atom. This is ensured by the superconducting gap in the QP spectrum, which keeps Shiba states with long excitation lifetimes and  decoupled from  other QPs (except indirectly via thermal effects and/or via Andreev processes~\cite{Ruby_2015}).  Moreover,   Shiba states are fairly unperturbed by the presence of close impurities. Our measurements, shown in the SI \cite{SOM}, reveal that two impurities separated by only 4 surface lattice parameters (1.4 nm) display  largely undisturbed Shiba conductance maps, what is attributed to the short localization length scale of Shiba states in three dimensions  \cite{Rusinov_1969,Gerbold_2015}.  This suggests that coupling  impurity-induced bound states together into extended magnetic structures  \cite{Ji_2008,Nadj-Perge_2014,Ruby_2015b} requires impurity separations well in the sub-nm range. In these structures, identification of the characteristic orbital symmetries of impurity bound states is thus a unique fingerprint to unveil their (channel-specific)  magnetic ground state, their magnetic coupling to other nearby impurities, and to follow the formation of extended Shiba bands.


\section{Methods}
\subsection{STM measurements}
The experiments were conducted in a commercial SPECS GmbH Low-Temperature (1.2 K) STM, under Ultra-High Vaccum conditions. Pb(111)  films (thickness $>$ 100 nm) on SiC(0001) substrates  show crystalline grains with diameters larger than 300 nm. To increase the energy resolution, the STM tip was repeatedly embedded in the Pb film until a superconducting tip was obtained. The full superconducting state of the tip was proved by performing STS spectra on bare Pb regions and showing that the two  sharp coherence peaks appear at a distance close to 4$\times\Delta$, where $\Delta$=1.35 meV, as in Fig. 2a. Contrary to single-crystal measurements \cite{Ruby_2015c} the spectra on bare Pb films never showed the characteristic double gap structure, probably due to the small film thickness compared to the Pb coherence length-scale. Cr atoms were deposited on the Pb(111) surface at a surface temperature of $\sim$ 15K. 
The spectra shown were obtained using a lock-in amplifier, with modulation $V_{rms}$=10 $\mu$V at 938.5 Hz. The Shiba conductance maps were obtained from a matrix of 52$\times$52  dI/dV(V) spectra measured in a region of 4.8 nm$^2$ over the Cr impurity.     The corresponding topography images do not show any of the features observed in the Shiba maps (see inset of Fig. 2d).  Analysis of STM and STS data was performed with the WsXm \cite{wsxm} and SpectraFox \cite{RubySoftware} software packages. 

\subsection{Theoretical details}
Standard calculations using DFT were performed to reproduce the total
energy and the electronic structure of different configurations of a Cr
atom on Pb (111) (see SI \cite{SOM}).  The calculations unequivocally predict a
subsurface configuration under the surface  bridge site, with considerably
distortion of the surface Pb layer. A minimal Pb-Cr cluster is used
to determine the local electronic structure near the Cr atom. Using
the wavefunctions of this cluster, the Bogoliubov-de Gennes equations
are approximated, in the presence of an exchange field, obtaining the
equivalent of Rusinov's equations using generalized scattering channels
and a non-free-electron normal metal. The equations are
further approximated to yield the spatial distribution
of the Shiba states, as detailed in SI \cite{SOM}.

\section*{Acknowledgements}
We acknowledge fruitful discussions with Sebastian Bergeret, Katharina Franke, and Benjamin Heinrich.
DJC acknowledges the European Union for support under the H2020-MSCA-IF-2014 
Marie-Curie Individual Fellowship programme proposal number 654469. 
We acknowledge financial support from Spanish
MINECO (Grants No. MAT2015-66888-C3-2-R and MAT2013-46593-C6-1-P) and Diputacion Foral de Gipuzkoa for grant N$^\circ$ 64/15.


\bibliography{crpb}

\end{document}


\title{Mapping the orbital structure of impurity  bound states in a superconductor\\ Supplementary Material}

\author{D.-J. Choi}
\affiliation{CIC nanoGUNE, Tolosa
	Hiribidea 76, 20018 San Sebasti\'an - Donostia, Spain}

\author{C. Rubio-Verd\'u}
\affiliation{CIC nanoGUNE, Tolosa
	Hiribidea 76, 20018 San Sebasti\'an - Donostia, Spain}

\author{J. de Bruijckere}
\affiliation{CIC nanoGUNE, Tolosa
	Hiribidea 76, 20018 San Sebasti\'an - Donostia, Spain}
\affiliation{Kavli Institute of Nanoscience, Delft University of Technology, Lorentzweg 1, 2628 CJ Delft, The Netherlands}

\author{M.M. Ugeda}
\affiliation{CIC nanoGUNE, Tolosa
	Hiribidea 76, 20018 San Sebasti\'an - Donostia, Spain}\affiliation{ikerbasque,
	Basque Foundation for Science, Bilbao, Spain}

\author{N. Lorente}
\affiliation{Centro de F\'{\i}sica de Materiales CFM-CSIC, 20018 San Sebasti\'an - Donostia, Spain}
\affiliation{Donostia International Physics Center (DIPC), Paseo Manuel de Lardizabal 4, 20018 San Sebasti\'an-Donostia, Spain}

\author{J.I. Pascual}
\affiliation{CIC nanoGUNE, Tolosa
	Hiribidea 76, 20018 San Sebasti\'an - Donostia, Spain} \affiliation{ikerbasque,
	Basque Foundation for Science, Bilbao, Spain}

\maketitle

\section{The adsorption of C\lowercase{r} atoms on P\lowercase{b}(111)}

Chromium atoms were directly evaporated onto sample stage of our SPECS JT-STM, hosting the  Pb(111) film at $\sim$
15K . The STM topography of individual Cr atoms shows a distinctive elongated shape,
signaling a 2-fold symmetry of the adsorption site (Fig. S1a). The longitudinal axis of the elongated shapes points towards the three main crystallographic   axis of the Pb(111) surface (see
Figs.~S1c and S1d).  The spectroscopic features are identical for the three differently adsorbed Cr atoms, just with spatial shapes rotated 120$^o$. The height of Cr atom is relatively low, typically amounting $\sim 50\,$ pm, as for the atom shown in Fig. S1a (Fig.~S1b). Both, the low height and elongated shapes along crystallographic direction agrees with  DFT findings shown in the main text, and described  next.  

\begin{figure} [h]
\includegraphics[width=0.60\columnwidth]{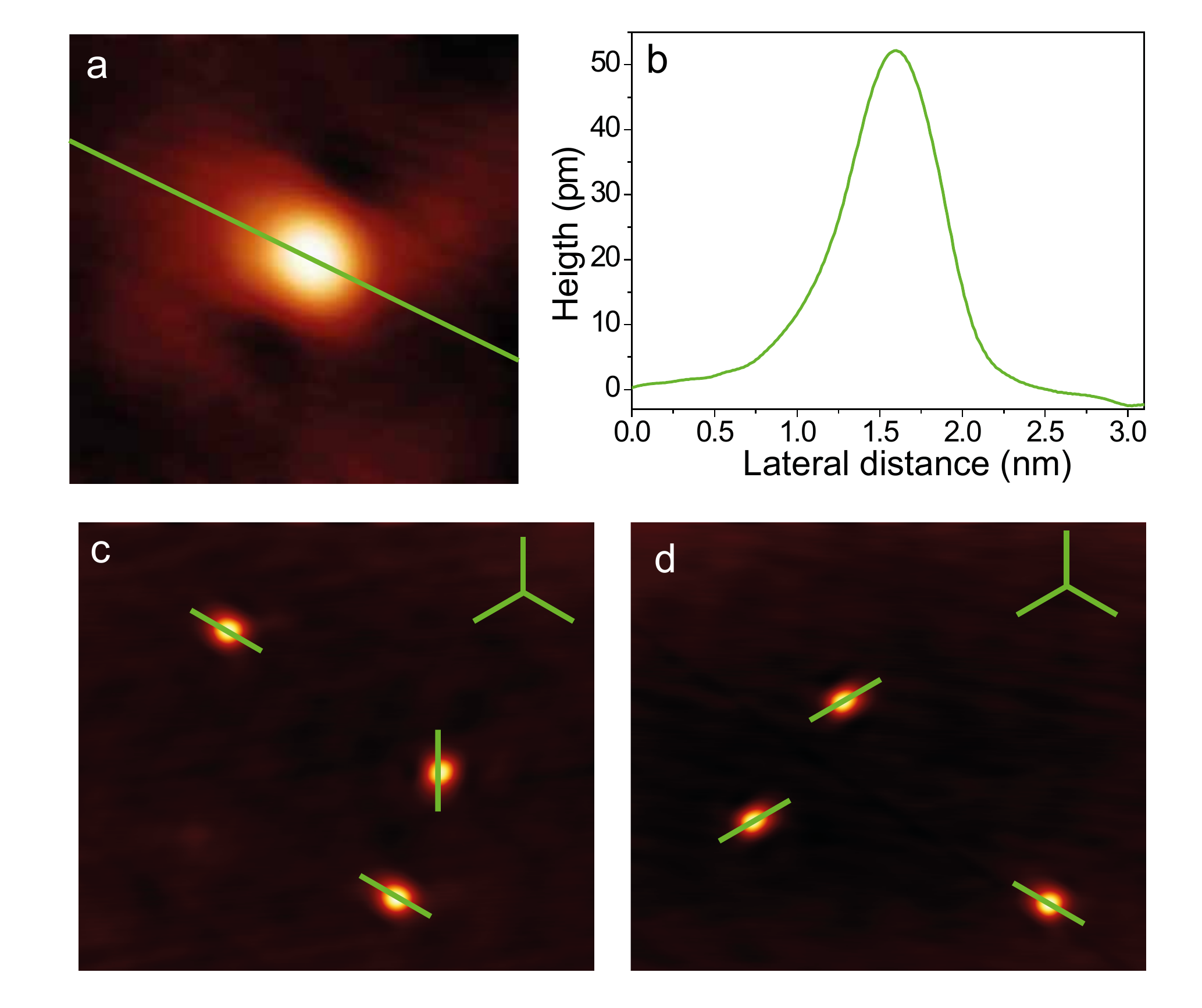}
	\caption{\label{FigureS1}
		(a) STM image of a Cr atom on the Pb(111) surface. They appear as elongated protrusions ($V$ = 4 mV, $I$ = 10 pA, size: $13\times13$ nm$^2$). (b) Height profile of Cr atom in (a). The height of the atom is $\sim 50\, pm$. (c) and (d) shows STM images of a $13\times13$ nm$^2$ area obtained with the same parameters. The elongation of Cr atoms points towards three directions separated by  120$^\circ$.
} 
\end{figure}
\newpage
\subsection{Density functional theory calculations}

The Density Functional Theory calculations  were performed using the  VASP~\cite{vasp} code. We simulated  the Cr/Pb(111) system with slab of 5 Pb layers   and a $5 \times 5$ Pb(111) surface unit cell.  We used the PBE exchange-and-correlation functional~\cite{PBE} with the
Dudarev \textit{et al}~\cite{dudarev_1998} correction for the intra-atomic
correlation of the Cr $d$-manifold, $U_{eff}=4$~eV, together with the PAW
method~\cite{PAW} and a cutoff energy of 250 eV. The k-point sampling
was $3\times 3\times 1$.  Forces were relaxed until they were smaller
than 0.02 eV/\AA.

We find a  lattice parameter for the Pb crystal of 5.02 \AA. This is 0.07 \AA~larger
than the experimental one as expected from the generalized-gradient
approximation PBE. We computed the  adsorption energy (energy gain with respect to having a Cr atom in the gas phase and a perfect surface) of a Cr atom at several high-symmetry sites on the Pb(111) surface and subsurface. The results are shown in Table~\ref{chem}. Subsurface adsorption is more favourable for both FCC-hollow sites and bridge sites, but less favourable for HCP-hollow sites   due to the presence of the  second-layer Pb atom. We computed next the potential barrier height   for a Cr atom to reach a sub-surface FCC hollow site, and found that this was only 21 meV. So, (hot)  Cr atoms deposited onto the Pb(111) surface  can easily cross this barrier and relax into a subsurface site before dissipating their thermal energy. A subsurface adsorption  agrees well  with the small experimental corrugation found in our experiment, as well as in
previous reports for the adsorption of Fe on Pb(111)~\cite{Ruby_2015}. 

\begin{table}[t]
	\centering
	\begin{tabular}{lcccc}
		& top & bridge & FCC & HCP \\
		\hline
		Surface      & $-0.812$ &  $-1.209$ & $-1.263$ &$-1.268$    \\
		
		Subsurface      & - & $-1.483$ & $-1.411$ & $-1.121$    \\
	\end{tabular} 
	\caption[] {\label{chem} 
		Adsorption energies in eV  of a Cr atom at different surface (atop the first Pb (111) layer),
		or subsurface (between the first and second layers) sites.   
		The subsurface bridge site corresponds to an adsorption configuration where the Cr atom is
		below the first layer but between the FCC and bridge sites.
	}
\end{table}

However, Cr atoms   show lower symmetry in STM images (two-fold, as seen in Fig. S1). Allowing full  relaxation of a Cr atom at an initial subsurface FCC site, we find that it slightly drift towards the bridge site,  finding its lower energy configuration (E$_{adsorp}$=-1.483 eV) at an intermediate point between subsurface bridge and FCC hollow sites. 
As shown in Figure 1 of the main text, in this configuration the  two closer bridge atoms on top are slightly displaced upwards, and the    Cr atom sits in the center of an (slightly deformed) octahedron of  Pb atoms approximately 3.2 \AA\
away. A seventh Pb atom of the surface
also lies within this distance, and the interaction of the Cr impurity with the Pb crystal    can be reduced to these 7 neighbouring Pb atoms.   This
configuration  exhibits  a  mirror plane, perpendicular to the (111) surface, in good agreement with the symmetry of the measured topography.

Figure~3c in the main text shows the (spin-polarized) density of states projected on the
$d$-manifold of the minimal cluster of Fig.~1, for a Cr atom embedded in
Pb (111) as described above.  Since this
cluster is a component of the Cr-Pb system, the electronic hybridization
between the cluster states with a large d-component (orbitals from 1
to 5 in Fig. 3) is  large, leading to broad occupied peaks in the PDOS. The corresponding $d$-derived states with  opposite spin polarization are unoccupied and lie above the vacuum level in the present calculation,
showing that the system is strongly spin polarized. Indeed, the above
magnetic moments corroborate this picture, Table~\ref{mag}.  The crystal
field around the Cr atom splits the d-manifold. We see a tendency to
degenerate two pairs of states as one would expect from an octahedral
symmetry, however the calculations show that the symmetry is lower due
to the partial reconstruction of the surface and the inclusion of a
seventh atom in the minimal cluster. As a consequence all degeneracies
are lifted and five different states are recovered. These gives rise to
five distinct Shiba states.

\begin{table}[h]
\centering
\begin{tabular}{lcccc}
               & top & bridge & FCC & HCP \\
        \hline
                Surface      & 5.1 &  4.9 & 4.9 & 4.9    \\

                Subsurface      & - & 4.1 & 4.0 & 4.1    \\
\end{tabular}   
\caption[] {\label{mag} 
Atomic magnetic moments in $\mu_B$ of the adsorbed Cr atom. On the
surface, the Cr atom largely keeps its nominal S=5/2 spin, but in between
the first and second Pb layers, the spin drops to S=2.  } 
\end{table}

Table~\ref{mag} shows the magnetic moments of the Cr atoms in the
above adsorption sites. As the coordination with Pb atoms increases
the magnetic moment decreases due to the hybridization of the Cr d-manifold with  the Pb bands.   On the surface,
the spin of the  Cr atom is 5/2.  The preferred subsurface adsorption
leads to a reduction of the Cr spin to S=2.

\section{Evaluation of Shiba state images}

The theory of Shiba states is well covered in the
literature~\cite{Yu_1965,Shiba_1968,Rusinov_1969b,Zarand_PRB_2008}.
Rusinov includes all ingredients in his excellent
article on superconductivity near a paramagnetic
impurity~\cite{Rusinov_1969b}. Here, we use his methodology,
extending it to the scattering states of the impurity as done by
Moca and co-workers~\cite{Zarand_PRB_2008}.  As explained by Balatsky
\textit{et al}~\cite{Balatsky_2006,Balatsky_PRB_2008} we can approximate
the STM image by $|u^\mu_\uparrow (\vec{r})|^2$ for positive bias and by
$|v^\mu_\downarrow (\vec{r})|^2$ for negative bias. Here, $u$ and $v$
are the quasiparticle amplitudes that enter the Bogoliubov-de Gennes
equations~\cite{Rusinov_1969b}.  Superscript $\mu$ denotes the scattering
channel. As remarked by Moca and co-workers~\cite{Zarand_PRB_2008}, we
can use the Kondo approximation written in terms of scattering channels
to treat the scattering of one electron off the paramagentic impurity.
We use the definition of eigenchannels given by Paulsson and
Brandbyge~\cite{Magnus_2007}. We can see that the eigenchannels will
diagonalize the hybridization function~\cite{Korytar_2011},
given by:
\begin{equation}
\Gamma_{m,m'}(\omega)=\sum_{n \vec{k}} V_{m,n \vec{k}} V_{n \vec{k}, m'} \delta (\omega-\epsilon_{n \vec{k}}).
\label{gamma}
\end{equation}
Here $V_{m,n \vec{k}}$ is the one-electron hybridization between the impurity's
orbitals, $m$ and the substrate's Bloch wave functions, given by $n, \vec{k}$, at
energy $\epsilon_{n \vec{k}}$.
By virtue of the Wolf-Schrieffer transformation, the eigenchannels also diagonalize
the potential and Kondo scattering interactions:
\begin{equation}
\hat{H}_{imp} = \sum_{\mu, \alpha} K^{\mu} \hat{\psi}^\dagger_{\mu,\alpha} (\vec{r}) \hat{\psi}_{\mu,\alpha} (\vec{r}) - \frac{1}{2}  \sum_{\mu, \alpha, \beta} J^{\mu} \hat{\psi}^\dagger_{\mu,\alpha} (\vec{r}) \vec{\sigma}_{\alpha,\beta} \cdot \vec{S}\hat{\psi}_{\mu,\beta} (\vec{r}).
\label{impurity}
\end{equation}
In this expression, $K$ and $J$ are the matrix elements of the
scattering potential and the exchange-coupling term, $\alpha,\beta$ are
spin-$\frac{1}{2}$ indices. Hence,
the above expression takes into account that the scattering eigenchannels
given by the field operators $\hat{\psi}_{\mu,\alpha}$ diagonalize
the interactions with respect to the scattering indices $\mu$ given by
the impurity's $d$-manifold.

The Shiba states can now be found by looking at solutions inside
the superconducting gap $\Delta$. The in-gap scattering prevents
spin-flips so that we can just take the non-spin-flip components
of Eq.~(\ref{impurity}). We adopt the Nambu notation following
Shiba~\cite{Shiba_1968} such that the impurity Hamiltonian acts on a
Nambu wavefunction, $\Psi(\vec{r})$, as:

\begin{equation}
\hat{H}_{imp} \Psi(\vec{r}) =  \sum_{\mu, \alpha, \beta}(K^{\mu}\rho_z \delta_{\alpha,\beta}-\frac{1}{2} J^{\mu} S \sigma_z \rho_z ) \psi_{\mu,\beta} (\vec{r}) \langle \psi_{\mu,\alpha} | \Psi \rangle .
\label{int}
\end{equation}
In this equation, tensorial products are not written and
$\sigma_z$ and $\rho_z$ are the $z$ Pauli matrices for the spin and
spatial sectors. The full Hamiltonian is divided in two, following
Rusinov~\cite{Rusinov_1969b}. The superconductor is treated in the BCS
model in the Nambu space. Hence, the Shiba states are solutions of:

\begin{equation}
[\hat{H}_{BCS} + \hat{H}_{imp}] \Psi(\vec{r}) = \epsilon \Psi(\vec{r})
\label{Schro}
\end{equation}
where $\epsilon$ is the energy of the Shiba state. We can solve the above equation using the BCS resolvent, $G_{BCS} (\vec{r}-\vec{r}\,')$, obtaining
\begin{equation}
\Psi(\vec{r}) = \int G_{BCS} (\vec{r}-\vec{r}\,') \hat{H}_{imp} \Psi (\vec{r}\,') d^3 r'.
\label{sol}
\end{equation}
The resolvent is given in terms of the normal-metal Bloch functions $\phi_{n \vec{k} \alpha}(\vec{r})$ by
\begin{equation}
 G_{BCS} (\vec{r}-\vec{r}\,')=\sum_{n \vec{k} \alpha} \frac{\phi_{n \vec{k} \alpha}(\vec{r}) \phi^*_{n \vec{k} \alpha}(\vec{r}\,')}{\epsilon^2-\xi^2_{n \vec{k}}-\Delta^2_n} [\epsilon+\xi_{n \vec{k}} \rho_z+\Delta_n \sigma_y \rho_y].
\label{resolv}
\end{equation}
Including the above equation in Eq.~(\ref{sol}) together with Eq.~(\ref{int}), we obtain the equivalent of Rusinov's
equations using generalized scattering channels and a non-free-electron normal metal:
\begin{eqnarray}
u^\mu_{\uparrow} (\vec{r})&=&\sum_{n, \vec{k}, \alpha} \frac{\phi_{n \vec{k} \alpha}(\vec{r}) }{\epsilon^2-\xi^2_{n \vec{k}}-\Delta^2_n}
[(\epsilon+\xi_{n \vec{k}})(K^\mu-\frac{1}{2}J^\mu S)\langle \phi_{n \vec{k} \alpha}|\psi_{\mu,\uparrow}\rangle
\langle \psi_{\mu,\uparrow}|u^\mu_{\uparrow} \rangle \nonumber \\
&+&\Delta_n(K^\mu+\frac{1}{2}J^\mu S) \langle \phi_{n \vec{k} \alpha}|\psi_{\mu,\downarrow}\rangle
\langle \psi_{\mu,\downarrow}|v^\mu_{\downarrow} \rangle]
\label{u1}
\end{eqnarray}
and for the occupied part:
\begin{eqnarray}
v^\mu_{\downarrow} (\vec{r})&=&-\sum_{n, \vec{k}, \alpha} \frac{\phi_{n \vec{k} \alpha}(\vec{r}) }{\epsilon^2-\xi^2_{n \vec{k}}-\Delta^2_n}
[(\epsilon-\xi_{n \vec{k}})(K^\mu+\frac{1}{2}J^\mu S)\langle \phi_{n \vec{k} \alpha}|\psi_{\mu,\downarrow}\rangle
\langle \psi_{\mu,\downarrow}|v^\mu_{\downarrow} \rangle \nonumber \\
&+&\Delta_n(K^\mu-\frac{1}{2}J^\mu S) \langle \phi_{n \vec{k} \alpha}|\psi_{\mu,\uparrow}\rangle
\langle \psi_{\mu,\uparrow}|u^\mu_{\uparrow} \rangle] .
\label{v1}
\end{eqnarray}

In order to evaluate the modulus square of the above two expressions
that give rise to the computed Shiba images, Fig.~3 of the main text, we
perform the following approximations. We assume we can replace the Bloch
functions by simple plane waves. This reduces the computational effort to
performing Fourier transforms over the first Brillouin zone, $\psi_{\mu}(\vec{k})=\langle \phi_{n \vec{k} \alpha}|\psi_{\mu}\rangle$.  Furthermore,
we assume that the scattering channels, $\psi_{\mu,\downarrow}$,
are independent of the spin as done in a restricted Hartree-Fock
approximation.  In this case, the above equations are simplified to

\begin{eqnarray}
u^\mu_{\uparrow} (\vec{r})&=& \frac{1}{\sqrt{vol}} \sum_{n,\vec{k}} \frac{e^{i \vec{k} \cdot \vec{r}} }{\epsilon^2-\xi^2_{n \vec{k}}-\Delta^2_n} \psi_{\mu}(\vec{k}) [(K^\mu-\frac{1}{2}J^\mu S)(\epsilon+\xi_{n \vec{k}}) \langle \psi_{\mu}|u^\mu_{\uparrow} \rangle
\nonumber \\
&+&\Delta_n(K^\mu+\frac{1}{2}J^\mu S) 
\langle \psi_{\mu}|v^\mu_{\downarrow} \rangle]
\label{u}
\end{eqnarray}
and for the hole part:
\begin{eqnarray}
v^\mu_{\downarrow} (\vec{r})&=&-\frac{1}{\sqrt{vol}} \sum_{n,\vec{k}} \frac{e^{i \vec{k} \cdot \vec{r}} }{\epsilon^2-\xi^2_{n \vec{k}}-\Delta^2_n} \psi_{\mu}(\vec{k}) [(\epsilon-\xi_{n \vec{k}})(K^\mu+\frac{1}{2}J^\mu S)
\langle \psi_{\mu}|v^\mu_{\downarrow} \rangle \nonumber \\
&+&\Delta_n(K^\mu-\frac{1}{2}J^\mu S) 
\langle \psi_{\mu}|u^\mu_{\uparrow} \rangle] .
\label{v2}
\end{eqnarray}
Despite the obvious simplification, these equations are still difficult
to solve. 
In order to sketch the shape of the resulting  Shiba states, we have focused on the spatial components of   $u^\mu_{\uparrow} (\vec{r})$ and $v^\mu_{\downarrow} (\vec{r})$. 
We assume that the overlaps of the scattering functions with
the quasiparticles coefficients are constant and we do not solve these
equations self-consistently. To estimate them, we use that a single eigenchannel enters the
quasiparticle functions, and then:
\begin{equation}
u^\mu_{\uparrow} (\vec{r})=\langle \vec{r} | \psi_{\mu} \rangle \langle \psi_{\mu} | u^\mu_{\uparrow} \rangle
\label{uexp}
\end{equation}
and from here we obtain that 
\[
\frac{u^\mu_{\uparrow} (0)}{v^\mu_{\downarrow} (0)} = \frac{\langle \psi_{\mu} | u^\mu_{\uparrow} \rangle}{\langle \psi_{\mu} | v^\mu_{\downarrow} \rangle}. 
\]
From the usual Rusinov's equations, Ref.~\cite{Rusinov_1969b}, we obtain
that this ratio is approximately 0.3 for typical coupling values. 
According to our DFT calculations the potential
scattering is much smaller than the Kondo one and we can approximate the states
by
\begin{equation}
u^\mu_{\uparrow} (\vec{r})\propto \sum_{n,\vec{k}} \frac{e^{i \vec{k} \cdot \vec{r}} }{\epsilon^2-\xi^2_{n \vec{k}}-\Delta^2_n} \psi_{\mu}(\vec{k}) [0.3 (\epsilon+\xi_{n \vec{k}}) 
+\Delta_n] 
\label{uc}
\end{equation}
and for the hole part:
\begin{equation}
v^\mu_{\downarrow} (\vec{r}) \propto \sum_{n,\vec{k}} \frac{e^{i \vec{k} \cdot \vec{r}} }{\epsilon^2-\xi^2_{n \vec{k}}-\Delta^2_n} \psi_{\mu}(\vec{k}) [\epsilon-\xi_{n \vec{k}}
- 0.3 \Delta_n]. 
\label{vc}
\end{equation}

We clearly see in these equations that the quasiparticle images are made from the
Fourier components of the scattering wave functions. We take as the scattering wave
functions the wavefunctions of each orbital depicted in Fig. 3 of the main text, as they  correctly reproduce
the impurity's states in the minimal cluster around the impurity.
We observe that  BCS factor multiplying the Fourier
components of the scattering wave functions, $1/(\epsilon^2-\xi^2_{n \vec{k}}-\Delta^2_n$), 
is basically a Dirac $\delta$ function centered about the Fermi
energy. Hence, it acts as a filter, selecting the Fourier
components for the wavevectors $\vec{k}$ of the Fermi surface, what \textit{de Facto} causes that only  band states $\xi_{n \vec{k}}$  within $\sim \Delta_n$ about the Fermi energy contribute to the quasiparticle amplitudes.

The spatial dependence comes from the contribution of the different  Fourier components, $\psi_{\mu}(\vec{k})$ evaluated at the Fermi sphere. 
The extended two Fermi surfaces of Pb assure that almost all  $\vec{k}$ components are well represented in the summations.  
The Fourier components of the scattering wavefunctions are further 
weighed by a factor proportional to 
\begin{equation}
0.3 (\epsilon+\xi_{n \vec{k}}) 
+\Delta_n 
\label{x1}
\end{equation}
for the empty-state quasiparticle wavefunction
and 
\begin{equation}
\epsilon-\xi_{n \vec{k}}
- 0.3 \Delta_n
\label{x2}
\end{equation}
for the occupied one. 

We can easily observe that the  coefficient of the hole component $v^\mu_{\downarrow}$, Eq.~(\ref{x2}), changes sign at the value
$\xi_{n \vec{k}}= \epsilon- 0.3 \Delta_n$, which is a positive number, smaller than $\Delta_n$.   
This means that, as the  $\vec{k}$  approaches the Fermi wavevector $\vec{k}_F$, the Pb band $\xi_{n \vec{k}}$ approaches the value at which the factor (\ref{x2}) changes sign, and the components  $\psi_{\mu} (\vec{k})$ undergo a phase-shift in the Fourier summation.  
However, for the particle component $u^\mu_{\uparrow}$, the change of sign of the factor in Eq.~(\ref{x1}) takes place at $\xi_{n\vec{k}}= -(\epsilon+\Delta_n/0.3)$, which is beyond the values weighted by the BCS factor in the summation  and, thus,  all $\psi_{\mu} (\vec{k})$ components contribute constructively with  roughly with the same coefficient. 

The consequences are that the empty states, $u^\mu_{\uparrow}$,
are basically proportional to the sum of $\psi_{\mu}(\vec{k}) e^{i
\vec{k} \cdot \vec{r}}$ over the Fermi surface because
there is no sign change of the coefficients for the allowed range of values. 
For Pb, the two Fermi
surfaces expand a quite complete subset of the first Brillouin zone,
leading to empty Shiba states $u^\mu_{\uparrow}$ maintaining same symmetries as
the scattering states, as shown in Figure~3d of the main text.
On the contrary, the filled states, $v^\mu_{\downarrow}$, suffer a change of sign  in the summation of Fourier components. The corresponding phase shifts cause a  considerably distortion of   the final symmetries. For simplicity, the maps of $|u^\mu_{\uparrow}|^2$ and $|v^\mu_{\downarrow}|^2$ in Fig.~3d are constructed employing a  simplified double-band Fermi surface covering most of the first Brillouin zone, but we expect that Fermi surfaces with pockets and large anisotropy in k-space will induce projected shapes of Shiba states very distorted from the originating impurity states. 


\section{Deconvolution of the tunnelling spectra}

We employed superconducting tips to increase the energy resolution of the tunnelling spectroscopy. 
For that reason, a double quasi-particle gap (4$\Delta$) is observed instead of one of width $2\Delta$ in the tunnelling spectra. In order to deconvolute the tunnelling spectra as if a normal-metal tip were employed, we used the following methodology \cite{Pillet_2010}. 

We assume that the current is proportional to the convolution of the tip's and sample's densities of states, $\rho_T (\omega)$ and $\rho_S (\omega)$, respectively. Hence, the differential conductance is given by
\begin{equation}
\frac{\partial I}{\partial V} \propto \int \rho_S (\omega) \frac{\partial\rho_T (\omega-eV)}{\partial V} [ f (\omega-eV,T)
-f (\omega,T) ] d \omega + \int \rho_S (\omega) \rho_T (\omega-eV)  \frac{\partial f (\omega-eV,T)}{\partial V} d \omega
\label{cond}
\end{equation}
where the Fermi distribution function is given by $f(\omega,T)$.

For the superconducting tip, we assume a BCS density of states (DOS)
with a phenomenological broadening parameter $\gamma$ which broadens
the sharp features at the gap edges~\cite{Dynes_1978}:
\begin{equation}
\label{eq:dos}
\rho_{T}(\omega) \propto \text{sgn}(\omega)\text{Re}\left(\frac{\omega}{\sqrt{\omega^{2}+2i\omega \gamma - \Delta^{2}}}\right).
\end{equation}
We estimate the values of $\Delta$ and $\gamma$ through simulation of an experimental spectrum on a bare region of the Pb(111) surface, i.e. away from
Cr atoms. We substitute Eq.~(\ref{eq:dos}) for both $\rho_S$ and $\rho_T$
in Eq.~(\ref{cond}) and find the parameters which show the best agreement
with the experimental empty tunnelling spectra. Good agreement is found
for $\Delta=1.341$ meV and $\gamma=0.0303$ meV at $T=1.1$ K. With the
approximate $\rho_{T}$ we can write down Eq.~(\ref{cond}) in discrete
form as a matrix operation on the (unknown) sample DOS:
\begin{equation}
\label{eq:conv}
\left[\frac{\partial I}{\partial V}\right] \propto \mathbf{C} [\rho_{S}]. 
\end{equation}  
where the left hand side is a column vector with the (known) experimental
spectrum, and $[\rho_{S}]$ is a column vector with the (unknown) DOS
of the sample. By comparing Eqs.~(\ref{cond}) and (\ref{eq:conv}), and
transforming the integral to a discrete sum over a finite energy range
we find the elements of the convolution matrix:
\begin{equation}
\mathbf{C}_{ij}=\frac{\partial \rho_{T} (\omega_{j}-eV_{i})}{\partial V}(f(\omega_{j}-eV_{i},T)-f(\omega_{j},T))+\rho_{T}(\omega_{j}-eV_{i})\frac{\partial f(\omega_{j}-eV_{i},T)}{\partial V},
\end{equation} 
in which $\omega_{j}$ are the discrete energy values and $V_{i}$ the bias voltage values of the tunnelling spectrum. In general, $\mathbf{C}$ has no inverse, meaning that we cannot solve for $\rho_{S}$ in Eq.~(\ref{eq:dos}). However, an approximate solution can be obtained by multiplying the left hand side of Eq.~(\ref{eq:conv}) by the Moore-Penrose pseudoinverse of $\mathbf{C}$. This procedure naturally yields the best approximate (least-squares) solution:
\begin{equation}
[\rho_{S}] \sim (\mathbf{C}^{*}\mathbf{C})^{-1}\mathbf{C}^{*} \left[\frac{\partial I}{\partial V}\right].
\end{equation}
In order to verify the accuracy of this procedure we re-convolute the obtained $\rho_{S}$ with $\rho_{T}$ through Eq.~(\ref{eq:conv}) and find excellent agreement with the original spectrum.

\section{Averaged tunnelling spectra}

\begin{figure} [t]
\includegraphics[width=0.7\columnwidth]{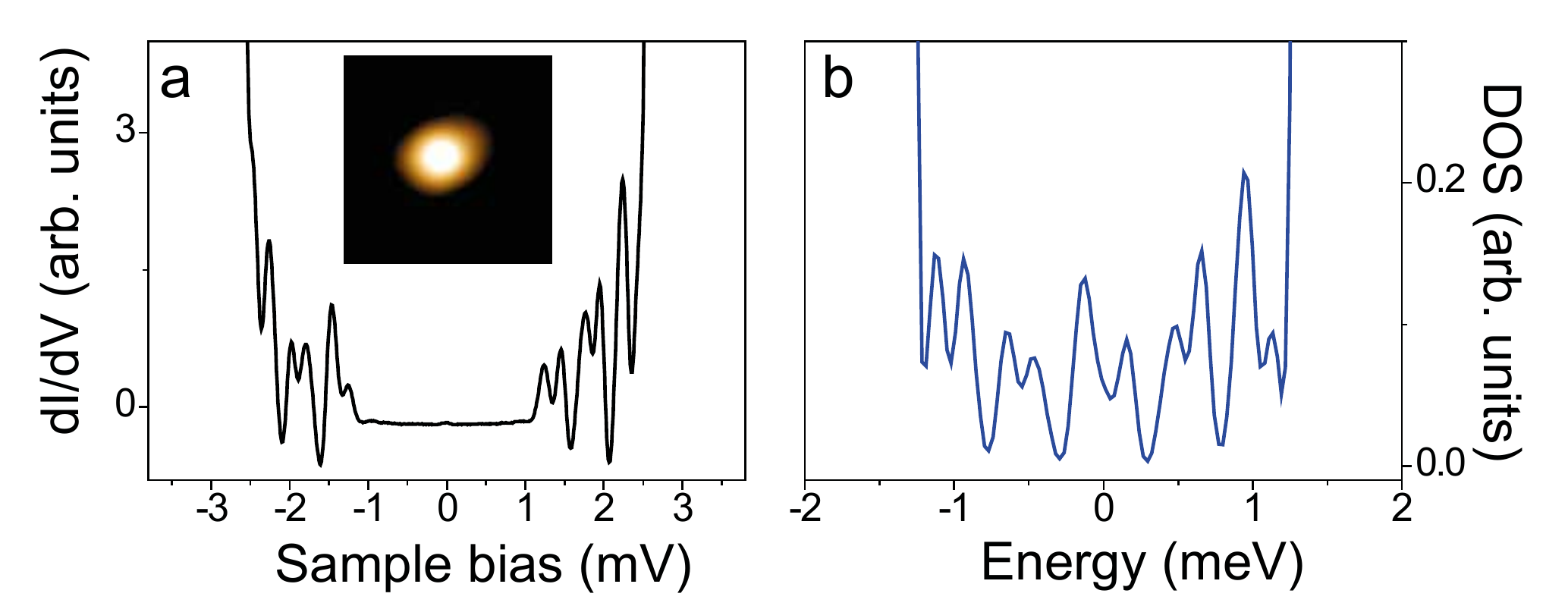}
	\caption{\label{FigureS3}
		$(a)$ Surface integrated tunnelling spectra obtained by summing all spectra  measured over the Cr atom   shown in the inset ($2.2\times2.2$ nm$^2$), and which used in Fig. 2d of the main text. $(b)$ Deconvoluted tunnelling spectra showing the five quasiparticle bound states.
		The peak intensity symmetry
		is partially restored. 
	}
\end{figure}
As we mention in the main text, due to the small   potential scattering term at the impurity, a large degree of particle-hole symmetry in the weight and amplitude of the Shiba states  is expected. But spectra shows usually very different peak intensities for each component of a Shiba pair, what is probably due to the inhomogeneous shapes of the states. To prove this, we summed over all the 52$\times$52 point spectra obtained over the $2.2\times2.2$-nm$^2$
area of Figure 2. The resulting spectrum, Fig.~\ref{FigureS3}, clearly  show that the surface-averaged intensities recover in part peaks with similar amplitude in their particle and hole components and, thus, the  source of asymmetry in a single spectrum is partly due to the different
the spatial distribution of each pair of functions 
\{$u_\uparrow (\vec{r})$, $v_\downarrow (\vec{r})$\}.  The deconvoluted spectra give five Shiba states at energies
$\epsilon_1$=1.1, $\epsilon_2$=0.9, $\epsilon_3$=0.62, $\epsilon_4$=0.45
and $\epsilon_5$=0.125 meV in good agreement with the energies obtained from
the spectra at a single spatial point.

\section{Conductance map of two neighbour C\lowercase{r} impurities}

\begin{figure} 
	\includegraphics[width=0.7\columnwidth]{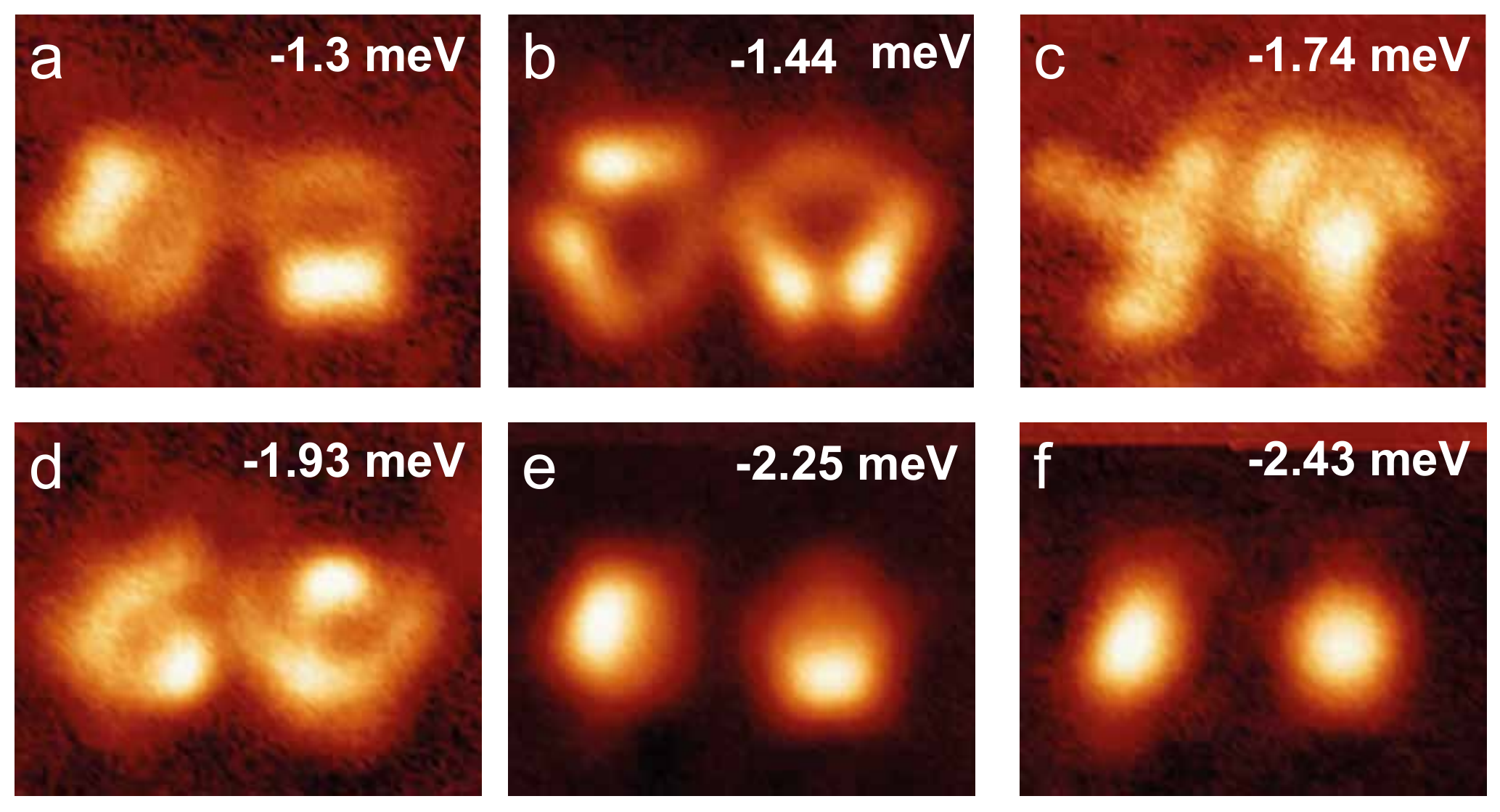}
	\caption{\label{FigureS4} 
		Conductance images over two  Cr impurities separated by 1.4 nm, mapping (b-f) the amplitude of negative Shiba states $\epsilon_n^h$ and (a) the thermal replica V$_6^h$. The maps are construted from a matrix of spectra over the shown area, as described in the main text. 
	}
\end{figure}

Figure~\ref{FigureS4} depicts the  conductance maps at negative Shiba quasiparticle states, performed as described in Figure~2 for the  single impurity case: a matrix of dI/dV spectra is acquired with stabilization voltage -4 mV and current 0.4 nA. The two Cr atoms are separated by four lattice unit cells ($\sim$~1.4~nm). The spatial distributions of the  Shiba states is similar to the isolated impurities, with some faint distortion in the shape of  states $\epsilon_3^h$ and $\epsilon_4^h$, but    rotated by~$120^\circ$, in agreement with  the three possible orientation of the adsorption site respect to the Pb(111) crystal directions. 
The bias at which these states are found agree within 0.05 mV with the
states of isolated impurities.  

The similarities, both energetically and spatially, with teh conductance maps of a single impurity in Fig. 2 lead us to the conclusion that impurities at a distance of $1.4$ nm exert no influence
onto each other.   Furthermore, this data show that mapping Shiba states offers the possibility of investigating channel specific magnetic coupling. 
\\

\bibliography{crpb_SOM}